\documentclass[aps,pra,twocolumn,amsmath,amssymb,superscriptaddress,showpacs]{revtex4-1}

\usepackage[utf8]{inputenc}
\usepackage{amsfonts}
\usepackage{amsmath}
\usepackage{amssymb}
\usepackage{amsthm}
\usepackage{fontenc}
\usepackage{graphicx}
\usepackage{xcolor}
\usepackage{textcomp}
\usepackage{epstopdf}
\usepackage{braket}
\usepackage{mathtools}
\usepackage{amsmath}
\usepackage{dcolumn}
\usepackage{multirow}
\usepackage{units}

\begin{document}
\newcommand{\abs}[1]{\lvert#1\rvert}
\title{  
Tuning transport properties on graphene multi-terminal structures by mechanical deformations}

\author{V. Torres}
\affiliation{Instituto de F\' isica, Universidade Federal Fluminense, Niter\' oi, Av.~Litor\^anea sn 24210-340, RJ-Brazil}
\affiliation{ MackGraphe, Mackenzie Presbyterian University, Rua da Consola\c c\~ ao 930,01302-907, SP-Brazil}
\author{D. Faria}
\affiliation{Instituto Polit\' ecnico, Universidade do Estado do Rio de Janeiro, Nova Friburgo, RJ, Brazil}
\author{A. Latg\' e}
\affiliation{Instituto de F\' isica, Universidade Federal Fluminense, Niter\' oi, Av.~Litor\^anea sn 24210-340, RJ-Brazil}
\email[A. Latg\'e]{andrea.latge@gmail.com}

\date{\today}

\begin{abstract}

Straintronic devices made of carbon-based materials have been pushed up due to the graphene high mechanical flexibility and the possibility of interesting changes in transport properties. Properly designed strained systems have been proposed to allow optimized transport responses that can be explored in experimental realizations. In multi-terminal systems, comparisons between schemes with different geometries are important to characterize the modifications introduced by mechanical deformations, specially if the deformations are localized at a central part of the system or extended in a large region. Then, in the present analysis, we study the strain effects on the transport properties of triangular and hexagonal graphene flakes, with zigzag and armchair edges, connected to three electronic terminals, formed by semi-infinite graphene nanoribbons. Using the Green's function formalism with circular renormalization schemes, and a single band tight-binding approximation, we find that resonant tunneling transport becomes relevant and is more affected by localized deformations in the hexagonal graphene flakes. Moreover, triangular systems with deformation extended to the leads, like longitudinal three-folded type, are shown as an interesting scenario for building nanoscale waveguides for electronic current.

\end{abstract}
%\pacs{ 73.63.-b, 72.25-b}

\maketitle

\section{Introduction}

The realization of mechanical strain on graphene structures is viewed as a promise route of tuning electronic and transport responses. Band-gap engineering based on mechanical deformation has been explored on graphene layers  and also on graphene nanoribbons\cite{Pereiragap,Torres2017,Lu2010, Sun2008}. Deformations from the ideal flat graphene sheet may appear naturally or can be intentionally induced by different experimental setups allowing modifications on the physical properties \cite{Evapillar}. Folded systems, for instance, are one of the possible products in the exfoliation process of graphene and have been theoretically investigated\cite{Rainis2011,Prada2010,Ramon2016,Naumis2017}. Bubbles and wrinkles are frequently observed in graphene structures grown by different techniques\cite{Lim2015,Tomori2011,Bao2009,Bunch2009,Incze2017}.  Deformations induced by electric fields acting on suspended graphene membranes are also shown to strongly affect their transport properties\cite{Fogler2008}.  Interesting symmetry features may be promoted by deformations in graphene. For example, the local density redistribution between distinct sublattices in graphene can be tuned by deforming the sample with a scanning tunneling microscope tip\cite{Alex2017}.  The emergence of pseudomagnetic fields that may reach quite high values\cite{Levy2010, Klimov, Lu2012} is a natural consequence when the mechanical perturbations are described as effective gauge field within continuum model descriptions\cite{Ando, Wakker2011, Daiara2013, Schneider2015,Juan2012,Stegmann2016}. Actually, a strong sublattice polarization in graphene system was proven to be a consequence of pseudomagnetic fields due to nonuniform strain distribution in graphene\cite{Schneider2015, Ramon2014, Salvador2013, Qi2014,Jauho2016}.  

In recent years, the field of valleytronics in graphene systems has been quite active. The major challenge being the proposal of new experimental setups able to produce a splitting in the transport of K and K' electrons, predicted by theoretical models\cite{Ramon2016,Ry2007,Chaves2010,Costa2015,Zenan2013,Dario,Ang2017,Asmar2017}. Deformed Hall bars have been proposed as proper devices to promote tunable polarized valley currents\cite{Peeters2017,Settnes2017}.  Triaxially  deformed graphene, with in- and out-of-plane strain are also addressed\cite{Mikkel2017} to induce valley-dependent electron trajectories  due to different pseudomagnetic polarization in the scattering region of the device. 

\begin{center}
\begin{figure}
\includegraphics[width=0.95\columnwidth]{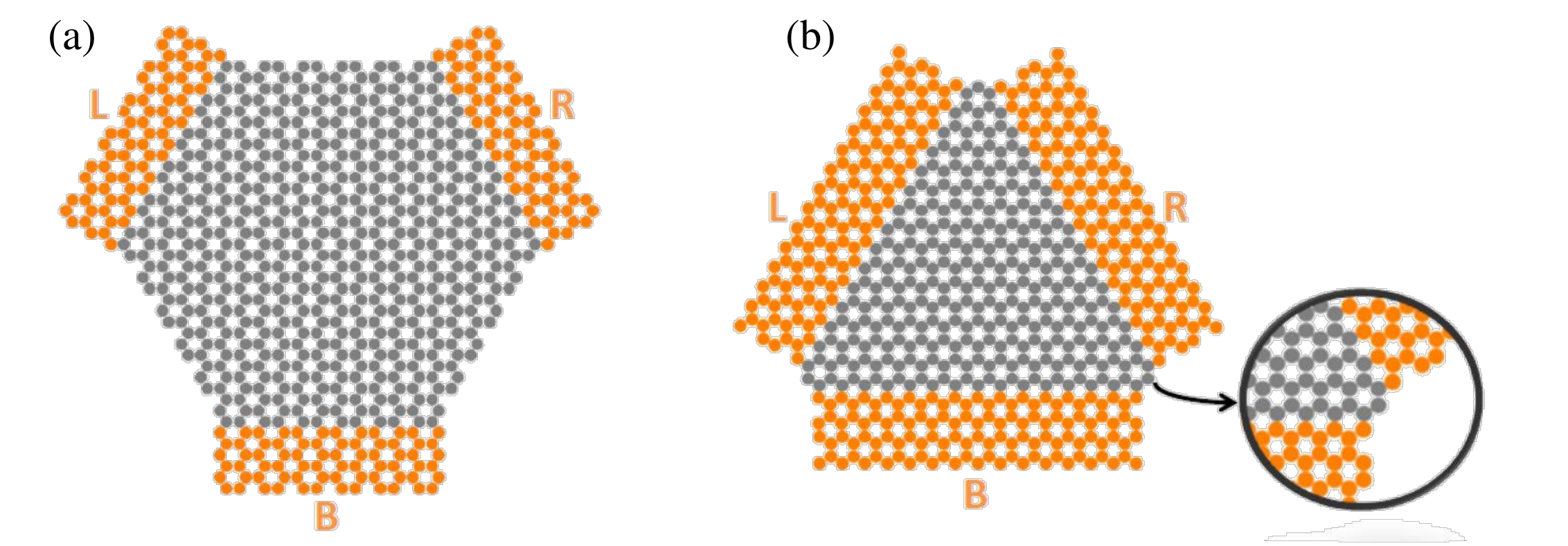}
\caption{(Color online) Schematic view of three-lead graphene quantum dots: (a) armchair-edge hexagonal flake and (b) triangular dot with zigzag edges. The leads are composed of semi-infinite zigzag and armchair nanoribbons, respectively. The zoom shows the details of a flake-lead corner.}
\label{device0}
\end{figure}
\end{center}
Confined systems like graphene quantum dots with different geometries have been largely explored such as the calculation of the electronic states of finite and unstrained hexagonal graphene flakes \cite{Heiskanen2008,Guclu2014,Sergey2011,Ezawa2008}.  It has been shown that there is a clear and scalable shell structure with the flake size in the case of armchair-edged hexagons, whereas for the zigzag hexagons, the level structure above the Fermi energy depends on the size of the hexagon. Similar studies have been reported for triangular graphene quantum dots, with discussions regarding the dependence of the energy spectra on the shape, edge, and sublattice symmetry of graphene quantum dots\cite{Guclu2014,Zarenia2011,Akola2008}. Notice that graphene equilateral triangular dots have been experimentally produced by etching processes on a graphene sheet\cite{Campos2009}. 

In addition,  a deformed graphene hexagonal flake has been studied taking into account a Gaussian shaped perturbation pinned at the center of the flake\cite{Peeters}. The sixfold symmetric wave function reported inside the Gaussian bump was discussed on the basis of the anisotropic effects of the strain along the main lattice direction. Quantum transport of in-plane triaxially strained triangular graphene quantum dots with three-terminals  has been discussed from atomistic mechanical simulations\cite{Zenan2013}, revealing that a quasi-uniform pseudomagnetic field induced by strain may restrict transport to Landau level and edge state-assisted resonant tunneling.  Undeformed three-terminal graphene systems have been proposed as nanoscale thermal valve and amplifiers\cite{Zhong2012}, due to the possibility of controlling the thermal transport in these systems.

\begin{center}
\begin{figure}[hbt]
\includegraphics[width=1\columnwidth]{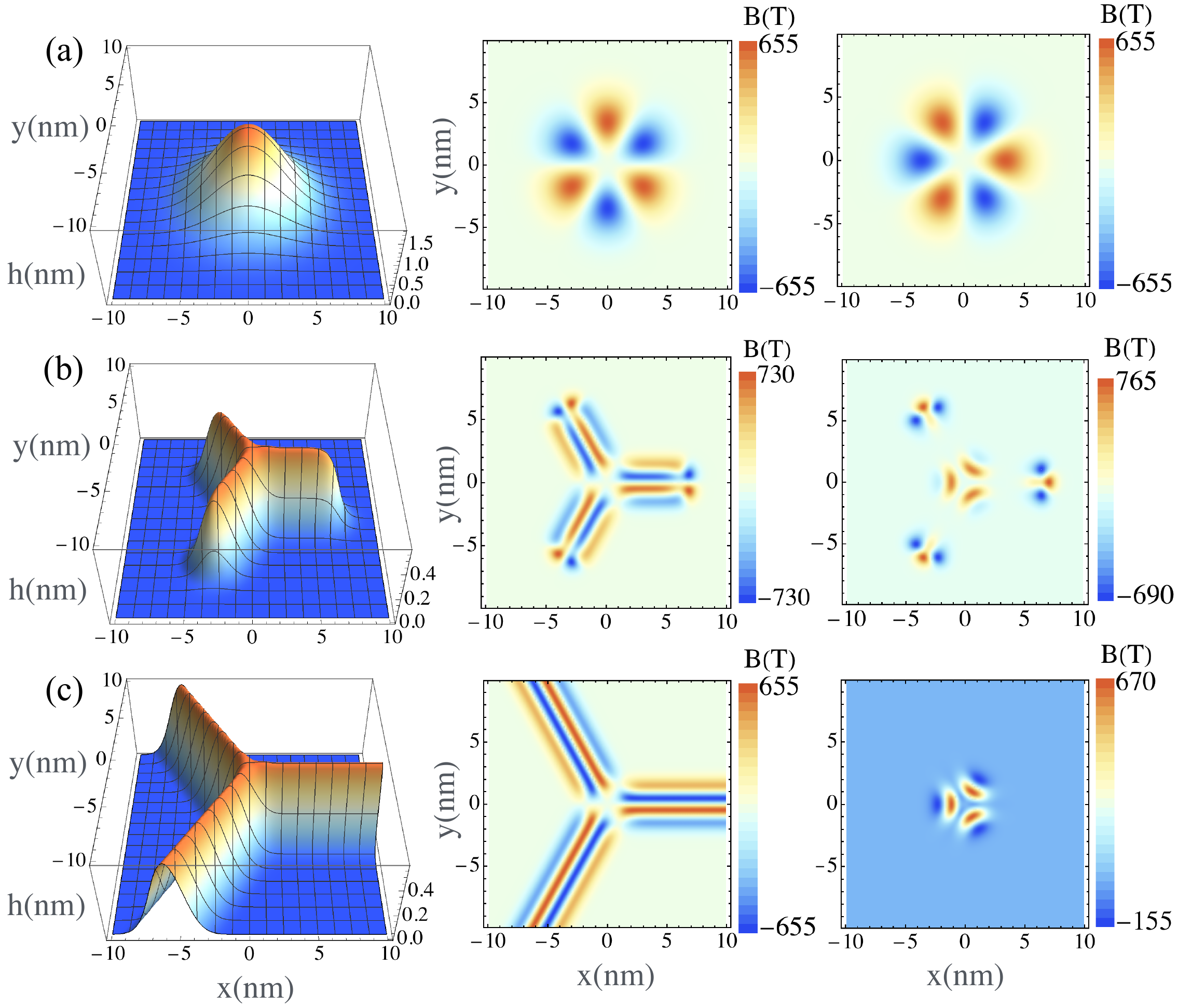}
\caption{(Color online) Spatial distribution of a 2D graphene sheet with (a) a Gaussian-like deformation ($A_g=12.5 a_{c}$, $b_g=28.0 a_c$, and $\alpha=20\%$), (b) a damped three fold-like strain and (c) an extended folded-like strain, both  with $A_f=4.0 a_{c}$, $b_f=10.0 a_{c}$, and $\alpha=16\%$. The corresponding pseudomagnetic fields are at the center and at the right panels in the colored maps, given in Tesla.  %Maximum values of B are: (a) 655 T, (b) 730 T, and (c) 655 T. 
The graphene zigzag (armchair) direction is parallel to the x-axis in the middle (right) pannels.}
\label{device}
\end{figure}
\end{center}

Motivated by the increasing experimental facilities in properly designing graphene systems, we propose an investigation on transport of a multi-terminal device composed of a graphene flake with hexagonal and triangular shapes. Three leads are connected, following seamlessly the graphene configuration of the central conductor, as sketched in Fig.\ \ref{device0}(a) for a hexagonal dot with armchair edges and in Fig.\ \ref{device0}(b) for triangular flake with zigzag edges.  Since we are interested in the coupling between mechanical and electronic transport properties in graphene multi-terminal systems, we consider Gaussian and fold-like deformations at the central conducting graphene flake. Differently from the case of in-plane triaxial strain, these deformations generate high anisotropic pseudomagnetic fields as shown in Fig.\ \ref{device}(a) and (b) for the corresponding deformations in 2D graphene sheets. We also consider the case in which the folds extend along three directions, as shown in Fig.\ \ref{device}(c). Notice that  in Fig.\ \ref{device}, the zigzag and the armchair direction are taken parallel to the x-axis in the panels at the middle and at the right, respectively. Our main goal is to find out the appropriate set up made of multi-terminal graphene flakes able to maximize the desired effects of strain on the transport responses. We analyze the pseudomagnetic fields induced by these strain models together with the edge effects. We calculate the electronic density of states (DOS) and the conductance of such flakes and local physical properties such as local density of states (LDOS) and current density in order to identify the occurrence of enhancement and suppressions of transport due to the deformation effects. We also analyze the possible use of deformed multi-terminal systems as  waveguides for electronic currents. 

%%%%%%%%%%%%%%%%%%%%%%%%%%%%%%%%%%%%%%%%
%%%%%%%%%%%%%%%%%%%%%%%%%%%%%%%%%%

\section{Structures and model}

The multi-terminal graphene system is composed of a hexagonal or a triangular graphene flake perfectly connected to three leads (see Figs. \ref{device0}(a) and (b)). The central conductors with armchair edges are coupled to zigzag leads and vice-versa. The leads are made of semi-infinite graphene nanoribbons of width given by $L^{A}=a_{c}(3N_{A}-2)/2$ for triangular and hexagonal armchair-edged flakes. For the case of zigzag edges, the widths are given by $L^{Z} = a_{c}\sqrt{3}[N_{Z}-1]$ for both triangular and hexagonal geometries.  Note that $N_{A,Z}$ is the number of outermost atoms in the edges of the hexagonal and triangular structures, for flakes with armchair and zigzag edges, respectively. The full system is described by a first-neighbor tight binding Hamiltonian (single band) given by,

\begin{equation} 
H =\sum_{\substack {<l,m>\\\sigma}} \gamma_{l,m} c_{l\sigma}^\dagger  c_{m\sigma} + \sum_{\alpha=1}^{3}[\sum_{\substack {<i,j>\\\sigma}} \gamma_{i,j}^{\alpha} c_{i\sigma}^{\dagger\alpha}  c_{j\sigma}^{\alpha}+\sum_{\substack {<i,l>\\\sigma}}h_{i,l,\sigma\begin{scriptsize}
{\footnotesize •\begin{small}
{\normalsize •\begin{large}
\end{large}}
\end{small}}
\end{scriptsize}}^{\alpha}] \,\,,
\label{HR} 
\end{equation}
where the first term describes the central deformed region, the second refers to the three terminals (deformed or not), with $h_{i,l}^{\alpha}$ being the coupling Hamiltonian that connects the central part to the leads. In the absence of deformation, the first neighbor hopping energy  $\gamma_{ij}$ is constant and given by \cite{parameter}  $\gamma_{0}=-2.75eV$.  
From here on we neglect the spin index $\sigma$ since we consider degenerate
solutions.  As the strain is turned on, the deformation changes the hopping as $\gamma_{ij} = \gamma_{0}e^{-\beta \left(\frac{l_{ij}}{a_c}-1\right)}$, where the parameter $\beta=\left|\frac{\partial\log \gamma_o}{\partial\log a_c}\right|=3.37$ for carbon-carbon bonds, with $a_{c}=1.42\AA$ being the interatomic distance in unstrained graphene. 
The new bond distance under strain  $l_{ij}=\frac{1}{a_c}\left(a_c^{2}+\varepsilon_{xx}x_{ij}^2+\varepsilon_{yy}y_{ij}^2+2\varepsilon_{xy}x_{ij}y_{ij}
\right)$ is given by the strain tensor $\varepsilon_{\mu\nu}=\frac{1}{2}\left(\partial_\nu u_\mu+\partial_\mu u_\nu+\partial_\mu h \partial_\nu h\right)$,
defined in terms of the in- and out-of-plane deformation, $u_\nu$ and $h$, respectively \cite{Landau}. 

We consider three types of out-of-plane deformations: a centro-symmetric Gaussian perturbation at the center of the flake\cite{Wakker2011,Daiara2013} and two fold-like configurations. 

The out-of-plane Gaussian deformation is written as,
\begin{equation}
h_g(x_i, y_i) = A_g e^{-[(x_i-x_0)^2+(y_i-y_0)^2]/b_g^2} \,\,,
\end{equation} 
where $A_{g}$ and $b_{g}$ describe the amplitude and width of the Gaussian strain, respectively, and the coordinates $(x_0,y_0)$ denote the central position of the strain that is pinned at the geometrical center of the flake, coinciding with the center of the hexagon and the triangle.

The other two fold-like deformations are modelled by the superposition of three angular dependent fold deformations\cite{Ramon2016,Torres2017}, with each fold deformation given by
\begin{eqnarray}
h_f(x_i, y_i,\phi) &=& A_f e^{-[(y_i-y'_0)cos(\phi)+(x_i-x_0)sin(\phi)]^2/b_f^2}\nonumber\\
&\times &F(x_i,y_i,L_s)\,\, ,
\end{eqnarray}
where $A_{f}$ and $b_{f}$ describe the amplitude and width of the strained fold, respectively, $y'_o$ defines the maximum position along the cross section of the fold distribution chosen to allow the maximum symmetric atomic configuration. The rotating angle $\phi$ of each fold is chosen to coincide with one of the three leads direction, $\phi=0$ and $\pm 2\pi/3$ for the three folds. Additionally, we can consider rotations of this deformation with respect to the initial angles as perturbation effects. $F(x_i,y_i,L_s)$ is a smoothed function used in the 3 paddles of the fold-like deformation. For both strained-fold configurations we chose a damping factor in the central part of the perturbation to guarantee the same maximum fold height at this region. 

In one of the fold-like cases, the maximum height is smoothed down from the center up to a particular fixed distance outlining the deformation range, defined here as $L_s$. As such, the deformation is localized at a central region, as depicted in left panel Fig. \ref{device}(b), more similar to the Gaussian bump perturbation, although the deformation symmetry is quite different. 

In the other fold-like configuration, the deformation is perfectly extended into the three leads coupled to the flake, what is hereafter named as all-folded device.  Notice that in the case where the deformations extend to the leads $F(x_i,y_i,L_s)=1$.  

The corresponding pseudomagnetic field spatial dependences \cite{Ando}, due to the hopping modification in the system, are shown in Fig.\ \ref{device}, for each deformation considered, and directions of the fold axis. It is important to notice that fold deformations with axial direction parallel to armchair edges (lines) do not produce pseudomagnetic fields\cite{Ramon2016,Torres2017} along these lines, except at the junction of the three folds.

The conductance is calculated in the Landauer approach within the Green's function formalism \cite{Dattalivro},
\begin{equation}
{G}^{LR}(\varepsilon) = \frac{2e^2}{h} Tr[\Gamma^{L}(\varepsilon) g^r(\varepsilon) \Gamma^{R}(\varepsilon) g^a(\varepsilon)]\,\,,
\label{GLR}
\end{equation}
where $g^{r(a)} $ is the retarded (advanced) Green's function of the conductor and $\Gamma^{L(R)}(\varepsilon)=i[\sum^{r}_{L(R)}(\varepsilon) - \sum^{a}_{L(R)}(\varepsilon)]$ is written in terms of the left (right) lead self-energies $\Sigma^{a}_{L(R)}$. 

In what follows, we analyze the density of states and conductance results for hexagonal  and triangular  quantum dots with armchair (Fig. \ref{Armchair}) and zigzag (Fig. \ref{Zigzag}) edges, connected to idealized leads composed of semi-infinite zigzag and armchair nanoribbon, respectively. In the sequence, we compare the effects of different mechanical deformations (Gaussian and fold-like) on the conductance results of the nanostructured systems. 

\section{Results}

\subsection{Hexagonal and Triangular flakes without strain}

\begin{center}
\begin{figure}[hbt]
\scalebox{0.33}{\includegraphics{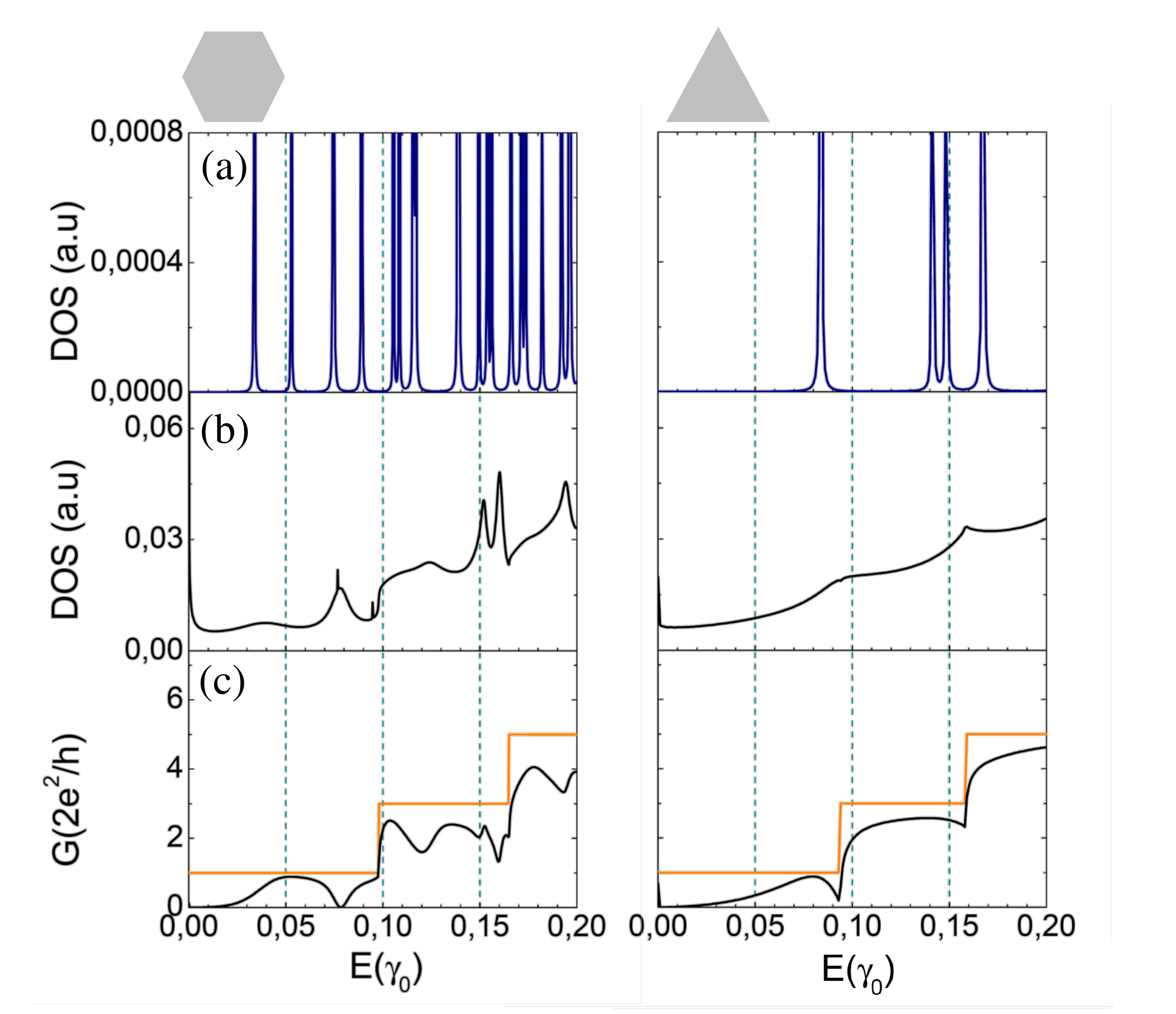}}
\caption{(Color online) Left and right panels: results for hexagonal $(N_{A}=50)$ and triangular flakes ($N_{A}=48$), respectively, with  armchair edges and connected to zigzag leads. (a) DOS for the corresponding isolated flakes. (b) DOS and (c) conductance for the full multi-terminal systems. The orange lines show the conductance of perfect zigzag nanoribbons, for comparison. {\label{Armchair} }}
\end{figure}
\end{center}

\begin{center}
\begin{figure}[hbt]
\scalebox{0.33}{\includegraphics{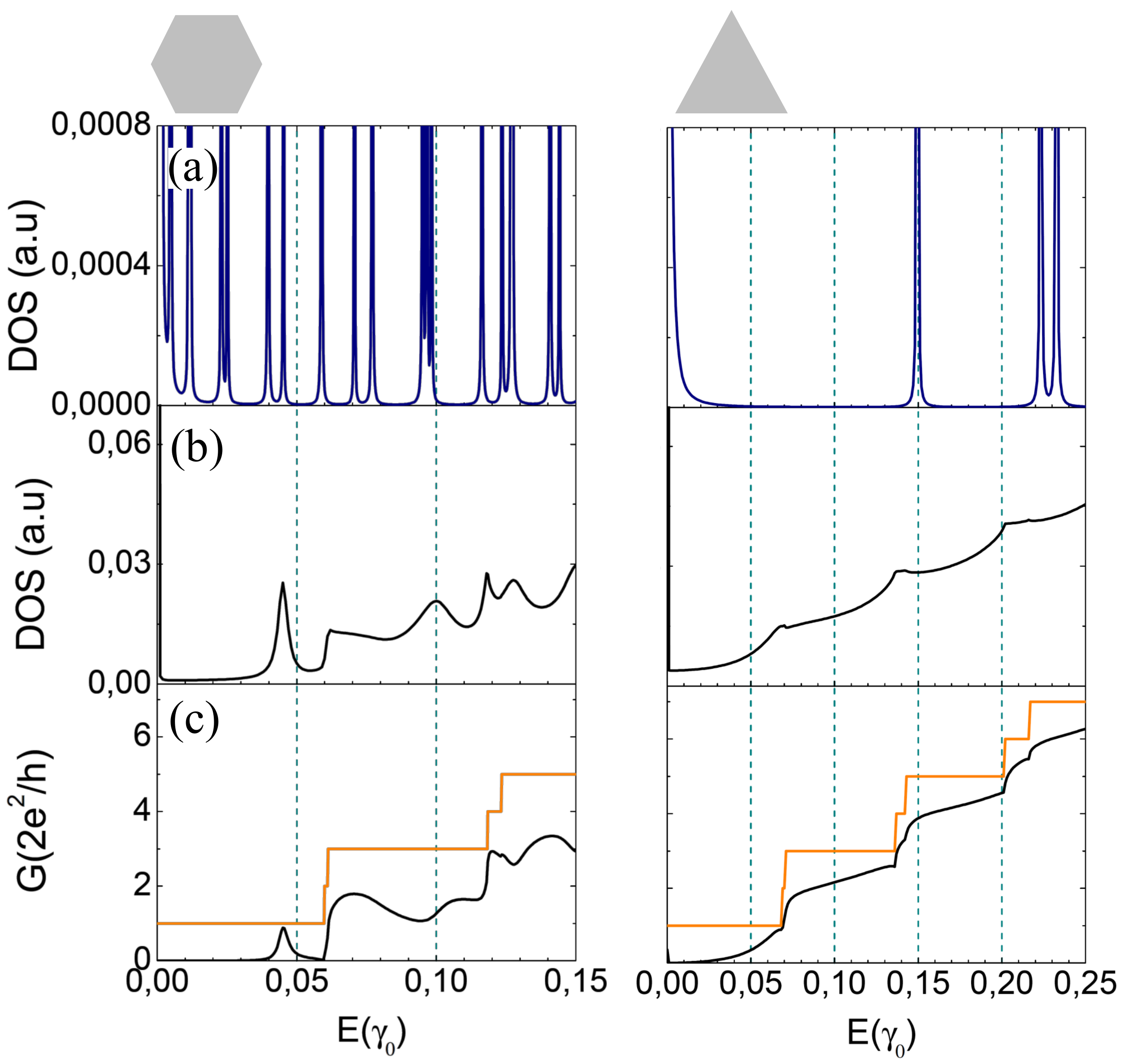}}
\caption{(Color online) Left panel and right panels: results for hexagonal $(N_{Z}=45)$ and triangular ($N_{Z}=39$) flakes, respectively, with zigzag edges and connected to armchair leads. (a) DOS for the isolated flakes. (b) DOS and (c) conductance for the full three-terminal systems. The orange lines show the conductance of perfect armchair nanoribbon, for comparison. {\label{Zigzag} }}
\end{figure}
\end{center}

To understand first the role played by the geometry on the electronic properties of typical graphene flakes, we start by discussing the density of states and conductance results for unstrained hexagonal and triangular-shaped quantum dots with armchair and zigzag edges, in Fig. \ref{Armchair} and Fig. \ref{Zigzag}, respectively. The results for hexagonal and triangular flakes are displayed on the left and right part of both figures. The correspondent energy levels for the case of isolated flakes are shown in the top parts of Fig.\ \ref{Armchair}(a) and \ref{Zigzag}(a). The results are in agreement with the fact that triangular armchair edge-shaped graphene nanoflake is a semiconductor (no zero-energy states) and that there are degenerated zero-energy states in all trigonal flake in the zigzag configuration\cite{Ezawa2008}. The conductance of pristine graphene nanoribbons of the same width as the leads are also shown with orange lines in part (c) of both figures, for comparison. Due to the particle-hole symmetry of the systems, the data are shown only in the positive energy range. 

The hexagonal and triangular flakes are chosen considering leads of similar widths. The undeformed armchair flakes connected to  zigzag terminals are formed by $N_{A}=50$ and $N_{A}=48$ for hexagonal and triangular flakes, respectively (Fig. \ref{Armchair}). In the case of zigzag-edged flakes $N_{Z}=45$ and $N_{Z}=39$ for hexagonal and triangular dots, respectively (Fig.\ \ref{Zigzag}). As a general result, we notice that for both edge configurations (Figs.\ \ref{Armchair} and \ref{Zigzag}), 
 the number of carbon atoms layering in the central part of hexagonal flakes is much greater than the number found in the corresponding triangular quantum dot case. This follows straightforward since the hexagon area is six times greater than the area of a triangle with the same side size. As a consequence, the number of energy levels of the isolated dot contained in the first conducting channel energy range is quite superior for the case of the hexagonal flakes as compared to the triangular quantum dots, for both armchair and zigzag edge configurations. 

%%%%%%%%%%%%%%%%%%%%%%%%%%%%%%%%%%%%%%%%%%%%%%%%%%%%%%%%%%%%%%%%%
%%%%%%%%%%%%%%%%%%%%%%%%%%%%%%%%%%%%%%%%%%%%%%%%%%%%%%%%%%%%%%%%%

For the hexagonal structures, a six-fold to three-fold symmetry reduction occurs when the central conducting flake is coupled to terminals, and only those states that are compatible with both symmetries are maintained. Some broad resonances and anti-resonances are formed, as can be seen in the left panels of Figs. \ref{Armchair} and \,\ref{Zigzag}.  In the low energy range swept by the first plateau, the results for the armchair-edged flake  show sharp DOS peaks coinciding with minima in the conductance.  At zero energy, this behavior can be observed for both zigzag or armchair flakes, thus devising a semiconducting system, even when the leads are  metallic. Other small sharp peaks, that do not contribute for transport, are observed in the case of the hexagonal armchair-edged flake, in left panel in Fig.\ \ref{Armchair}(b) and (c). These states are consequence of destructive interference between states from the disconnected central conducting flake and the nanoribbons leads\cite{BICs,Daiara2015}.  The first broaden DOS peak formed by conducting states can be associated with the first resonant mode in each hexagonal flake\cite{Zenan2013}.

%%%%%%%%%%%%%%%%%%%%%%%%%%%%%%%%%%%%%%%%%%%%%%%%%%%%%%%%%%%%%%%%%
%%%%%%%%%%%%%%%%%%%%%%%%%%%%%%%%%%%%%%%%%%%%%%%%%%%%%%%%%%%%%%%%%

It is interesting to note the correspondence between the positions of the energy levels of the isolated armchair triangular flake (decoupled from the leads), found in Fig.\ \ref{Armchair}(a) (right panel), and the energy positions at which the number of transport channels changes in the conductance results (see the steps between plateaux) shown in Fig.\ \ref{Armchair}(c). 
 When the isolated graphene triangle is coupled to the leads, differently from the hexagonal example, no symmetry reduction happens in the coupled system. The resulting conductance of the multi-terminal structures  is similar to the nanoribbon step-wise conductance, although a small reduction of transport is observed, specially at the first plateau that is mostly suppressed. Notice also the presence of pronounced local minima at the "step-to-step" transitions, which have been observed in other systems, mainly when disorder is included, and are associated with interband mixing favored by the presence of perturbations\cite{Bagwell1990}.   

As in the case of  multi-terminals with armchair central flakes, when the zigzag-edged flake is connected to the leads, there is no transport at zero energy, forming semiconducting system, even with metallic leads [Fig.\ \ref{Zigzag}(b) and (c)], for both hexagonal and triangular geometries. The deviations from the step-like behavior of the hexagonal and triangle-shaped dot conductances are clearly revealed  for the case of metallic armchair ribbons with the same width, being more pronounced for the hexagonal dot.  We conclude this section noting that even without distortion the studied flakes present electronic properties that depend on the details of the geometry of the central conductor and also on the graphene nanoribbons. At low energies the DOS of the triangular flake is more similar to the graphene DOS, due to absence of dangling bonds in the central flake presented in the hexagonal configuration.

In the next section we consider the effect of different deformations for armchair and zigzag-edged flakes. The Gaussian deformation, for both flakes, is associated with distinct orientations of the pseudo magnetic field space distributions with respect to the position of the lead\cite{Ramon2014}. The pseudomagnetic field, for fold-like deformations extending to the leads or not, are expected to be different depending on each flake edge configuration\cite{Ramon2016,Torres2017}. We investigate each case individually, as follows.

%.................................................................
\subsection{Deformation effects}

\subsubsection{Gaussian deformation}

The height profile of a graphene layer with a Gaussian strain, given by an amplitude of $12.52 a_{c}$ and a standard deviation $\sigma=b/\sqrt{2}$ with $b_g=28 a_{c}$ is shown in Fig. \ \ref{device} (a). This  centro-symmetric deformation induces the formation of a non-homogeneous pseudomagnetic field, depicted in the middle and right parts of Fig. \ref{device} (a), which resembles a 
six petal flower\cite{Ramon2014,Incze2017}. The pseudomagnetic field may attain quite high intensities depending on the deformation parameters. 
 
\begin{center}
\begin{figure}[hbt]
\scalebox{0.27}{\includegraphics{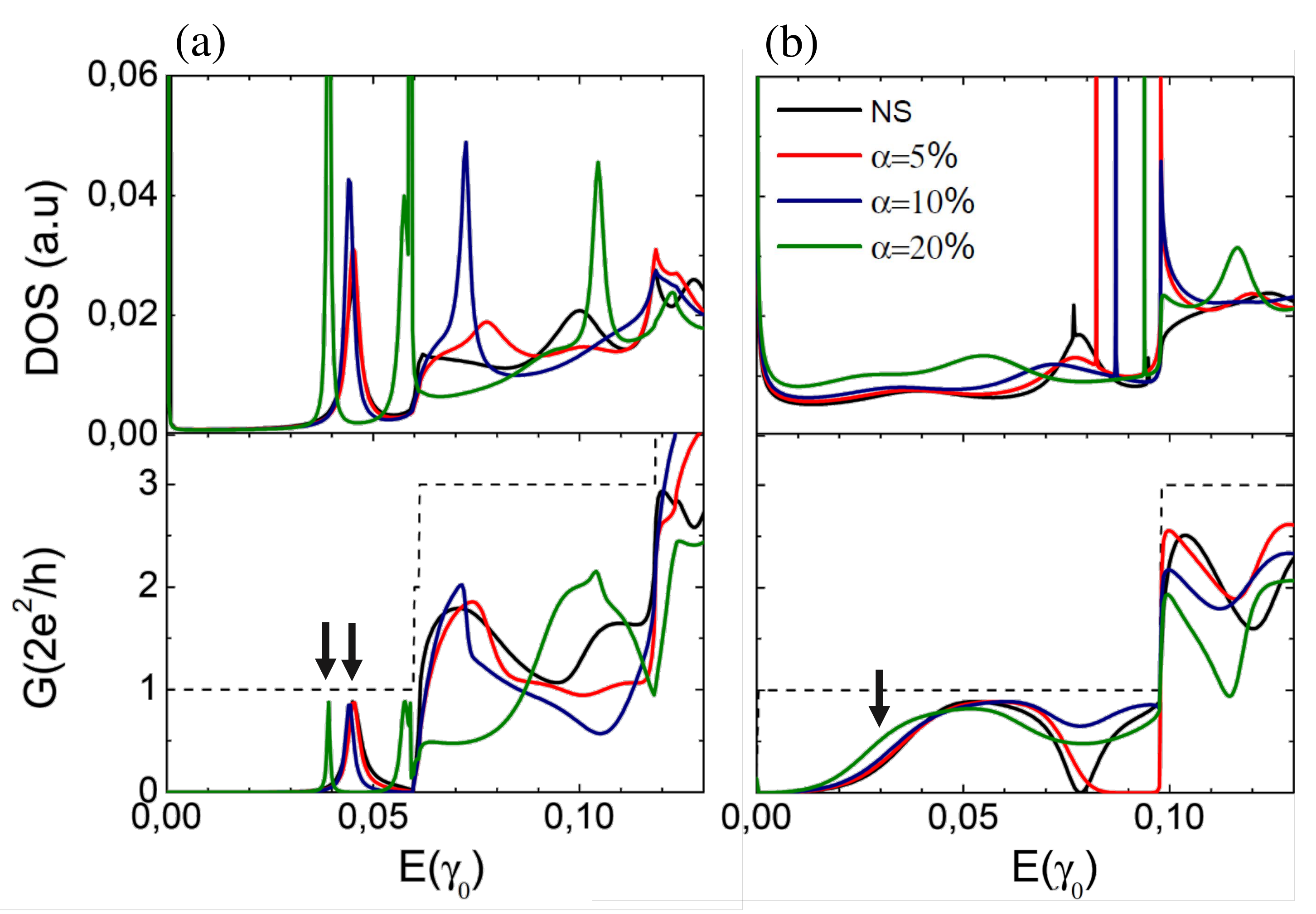}}
\caption{Electronic density of states (top panels) and conductance (bottom panels) for Gausssian deformed hexagonal graphene flakes with  (a) zigzag ($N_Z=45$) and (b) armchair ($N_A=50$) edges. The strain parameter is $b_g=28\;a_{c}$ and the dotted lines are the correspondent conductance of the undeformed leads.\label{hexgauss} }
\end{figure}
\end{center}

When the central hexagonal flake has zigzag edges (and armchair leads) the density of states results are more clearly affected. The Gaussian strain intensity is able to define a very narrow resonant state as shown in Fig. \ref{hexgauss}(a). One noticeable effect is the red shift of the resonance peak as the strain value $\alpha=(A/b)^2$ increases ($\alpha=0$ to $20\%$) together with a clear narrowing of the energy peak. Another resulting feature is the formation of new resonant states with the deformation, which are also shifted to low energies as $\alpha$ increases. A typical example is the resonance happening at the interface between the first and second conductance plateau of the corresponding pristine leads for the zigzag flake for $\alpha=20\%$. The resonant tunneling behavior as a consequence of applied strain has been observed for hexagonal zigzag graphene flakes, connected to multi-terminal, for triaxial deformations, with an intense constant pseudomagnetic field at the central part of the system\cite{Zenan2013}.  For this centro-symmetric deformation, the effect is not as strong because of the inhomogeneous magnetic field induced by the Gaussian deformation.  

In the case of the armchair-edged hexagonal flake, the Gaussian deformation does not disturb considerably the broaden states in the low energy range, as shown in the top part of Fig. 5(b). These states that contribute to conductance (see the bottom part of Fig. 5(b)) have a general trend of moving to low energies as $\alpha$ increases.  On the other hand,  localized states, in the energy range of the first conductance plateau for the isolated lead, shift to higher energies as alpha increases. The intensity of the peaks also increased  when the deformation is present. These states are related to the charge localization induced at the corners of the armchair-edged hexagonal flake, even without the deformation. When the system is deformed, it causes a charge redistribution in the system, and these states get more or less localized, depending on the interferences between them and other states of the continuum.  This feature is reflected in a complex evolution of the conductance.  We noticed that for $\alpha= 5\%$ a well defined gap is revealed in the conductance in the range of energy between $E\approx 0.08\gamma_0$ and $E\approx 0.1\gamma_0$, while for $5\% <\alpha<15\%$ the conductance increases as the strain increases. Finally, when $\alpha> 15\%$ the conductance decreases again in the same energy region. 

%%%%%%%%%%%%%%%%%%%%%%%%%%%%%%%%%%%%%%%%%%%%%%%%%%%%%%%%%%%%%%%%%
%%%%%%%%%%%%%%%%%%%%%%%%%%%%%%%%%%%%%%%%%%%%%%%%%%%%%%%%%%%%%%%%%

%%%%%%%%%%%%%%%%%%%%%%%%%%%%%%%%%%%%%%%%%%%%%%%%%%%%%%%%%%%%%%%%%
\begin{center}
\begin{figure}[hbt]
\scalebox{0.34}{\includegraphics{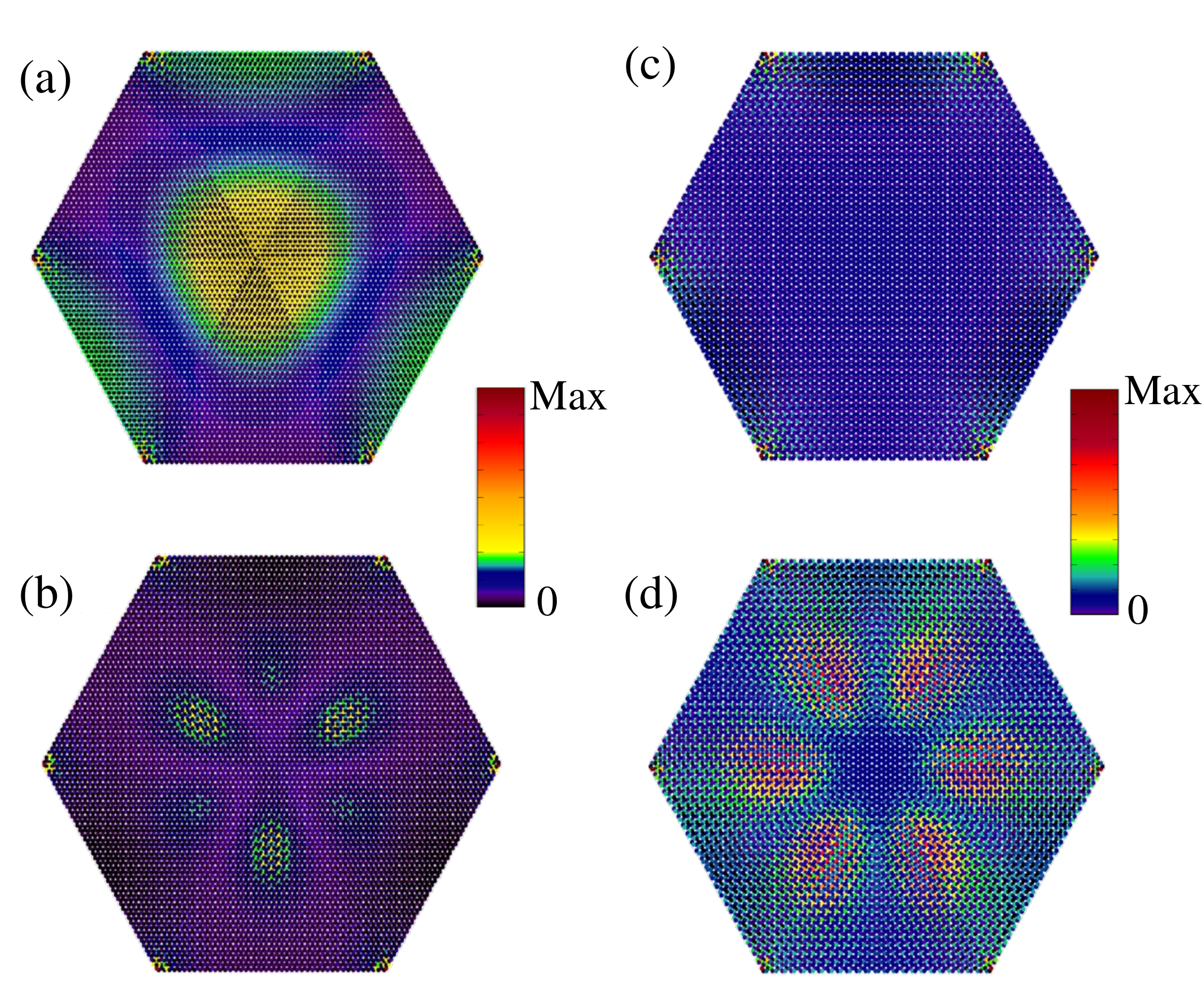}}
\caption{Local electronic density of states for undeformed [(a) and (c)] and Gaussian deformed [(b) and (d)]   hexagonal graphene flake systems. Hexagonal flake with  zigzag edges at (a) $E=0.045\;\gamma_0$ (undeformed) and (b) at $E=0.039\;\gamma_{0}$ (with strain). Armchair-edged system at  $E=0.03\;\gamma_0$ [(c) and (d)]. Strain parameter: $b_g=28\;a_{c}$ and
 $\alpha=20\%$. \label{ldoshg} }
\end{figure}
\end{center}
%%%%%%%%%%%%%%%%%%%%%%%%%%%%%%%%%%%%%%%%%%%%%%%%%%%%%%%%%%%%%%%%%

%We analyze the local density of states at the energy peaks marked with the arrows shown in the bottom part of Fig. \ref{hexgauss}.
Fig. \ref{ldoshg}(a) and (b) show the local  density of states at the first resonant state (marked with the arrows shown in the bottom part of Fig. \ref{hexgauss}) for the unstrained and for the  the Gaussian-like strained zigzag hexagonal flake, respectively.  In the unstrained case, we notice that the first resonant state ($E=0.0459\;\gamma_{0}$) is highly concentrated at the center of the flake and at the edges with no leads. For the strained system at $E=0.039\;\gamma_{0}$ (red-shifted peak) the center region exhibits a different pattern, typical of Gaussian-like pseudomagnetic fields \cite{Ramon2016}. Notice that both results exhibit three-fold symmetry due to the presence of the three leads and because of the symmetry of the deformation considered.  For armchair-edged dot, at $E=0.03\;\gamma_{0}$ (see Fig. \ref{hexgauss}(b)), the unstrained flake presents a homogeneous electronic distribution along the system, with highly localized states at the six corners. Under strain, these corner states are preserved but a "petal-like" pattern is observed within the dot, similar to the pseudo-magnetic field distribution formed in strained graphene system.

For the triangular configuration only a subtle decrease of the conductance was found in general, without introduction of new interesting features. We should say that the transport of graphene triangular flakes are almost transparent for such Gaussian deformations. The absence of symmetry break in the  triangular flake configuration and the fact of all the edges being saturated, differently from the hexagonal case, explain such distinct behavior.  

\subsubsection{Damped fold-like deformation }

\begin{center}
\begin{figure}[hbt]
\includegraphics[width=1.03\columnwidth]{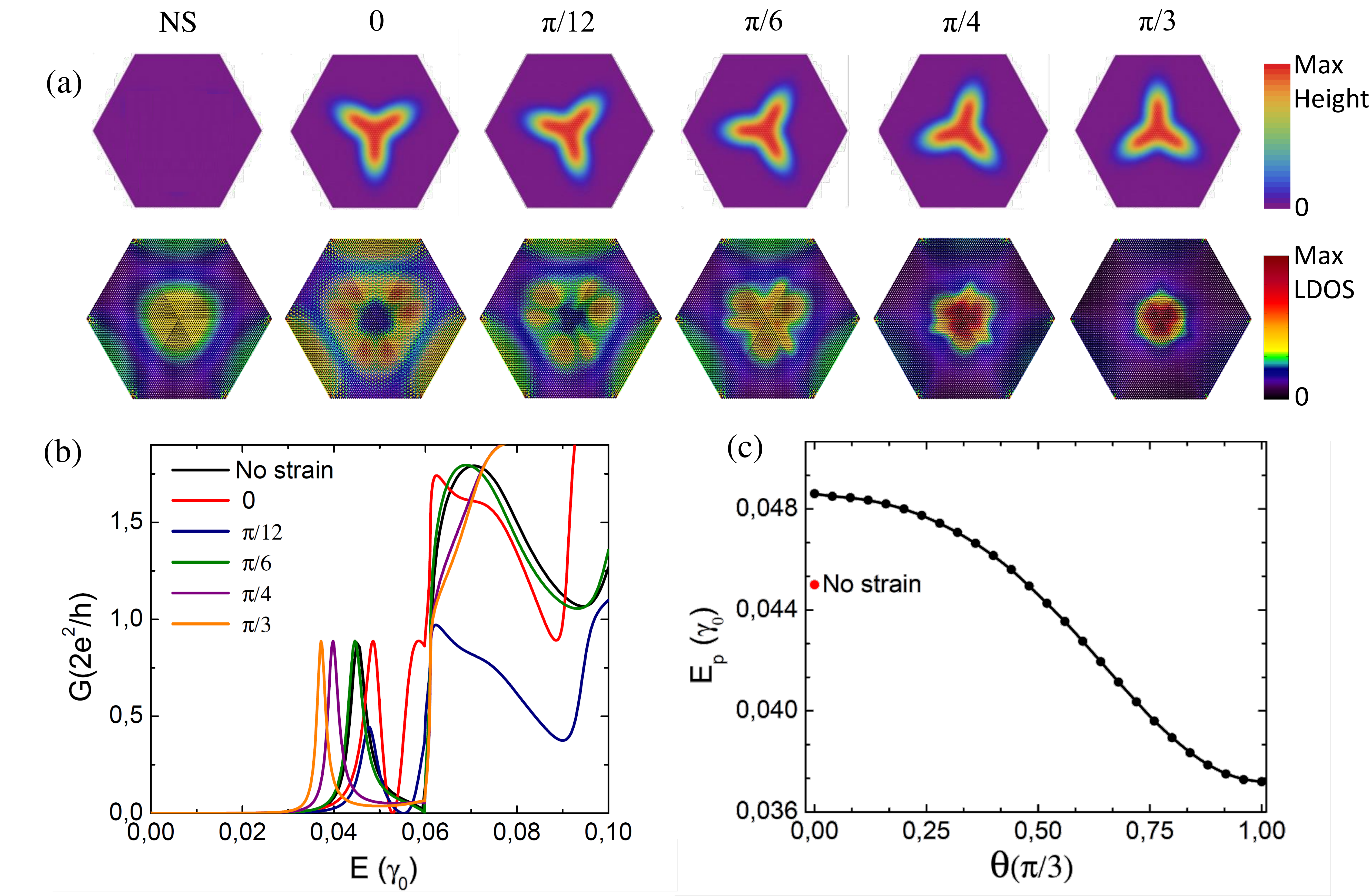}
\caption{(a) Spatial distribution (top) and LDOS at the resonant peak (bottom) of a zigzag hexagonal graphene flake ($N_Z=45$) with  smoothing folded deformation ($A_{sf}=4\;a_{c}$, $b_{sf}=10\;a_{c}$) for different values of the relative rotation angle with respect to the leads.  (b) Conductance results and (c) energy dependence of the resonant state on the rotation angle.  The first-energy state of the undeformed system is indicated with the red point. \label{foldamor} }
\end{figure}
\end{center}

For the case of the smoothed three-fold deformation, %smoothed out from the central part of the flake, being completely suppressed before arriving at the interfaces with the leads, 
we restricted our discussion to the configuration of a hexagonal graphene flake with zigzag edges. We analyze the possibility of tuning the zigzag hexagon resonant peak, by considering different rotation angles $\theta$ between the three-fold axis and the nanoribbon (lead) directions. 
The rotation angle must also be viewed as an attempt of introducing a kind of "disorder" into the folded graphene flake and, therefore, as a possibility of studying the transport problem within a more realistic scenario.
The case where $\theta=0$ is related to the geometrical configuration where the fold axis and the lead edges are parallel, as shown in the top panel of Fig. \ref{foldamor}(a), together with other rotation angle examples. Notice that the leads are formed by armchair nanoribbons and as such ($\theta=0$), the pseudomagnetic field profile generated by the folds is the one shown in the right panel in Fig.\ \ref{device} (b), for the same fold parameters. For $\theta=\pi/6$, the corresponding pseudomagnetic field profile is exhibited in the center panel of Fig.\ \ref{device} (b).  Fig. \ref{foldamor}(b) displays the conductance results for different rotation angles. The first resonant state of the undeformed system (black curve) is shifted to higher energy when the folded strain is considered at $\theta=0$,  and then, as $\theta$ increases, this resonant peak is moved to lower energies as shown in  Fig.\ \ref{foldamor}(b). The energy variation of the peak position as a function of the rotation angle is presented in Fig. \ref{foldamor}(c), corresponding to a range of $0.01\gamma_0$ ($\approx 30$meV). The energy position of the localized state for the undeformed system is marked in the figure by a red point for comparison.
Moreover, it is interesting to notice that for the particular value of $\theta$=$\pi/6$ the undeformed conductance results are almost recovered. The LDOS at the resonant peaks shown in the bottom part of Fig.\ \ref{foldamor}(a), for each one of the  rotation angle values, reveals the great similarity with the case of no strain: large concentration in the central region and same charge population along the hexagonal sides.  When $\theta$=$\pi/6$ the resonant peak is then just lightly affected.

%%%%%%%%%%%%%%%%%%%%%%%%%%%%%%%%%%%%%%%%%%%%%%%%%%%%%%%%%%%%%%%%%
%%%%%%%%%%%%%%%%%%%%%%%%%%%%%%%%%%%%%%%%%%%%%%%%%%%%%%%%%%%%%%%%%

\subsubsection{Extended fold-like deformation}

%%%%%%%%%%%%%%%%%%%%%%%%%%%%%%%%%%%%%%%%%%%%%%%%%%%%%%%%%%%%%%%%%
\begin{center}
\begin{figure}[hbt]
\scalebox{0.28}{\includegraphics{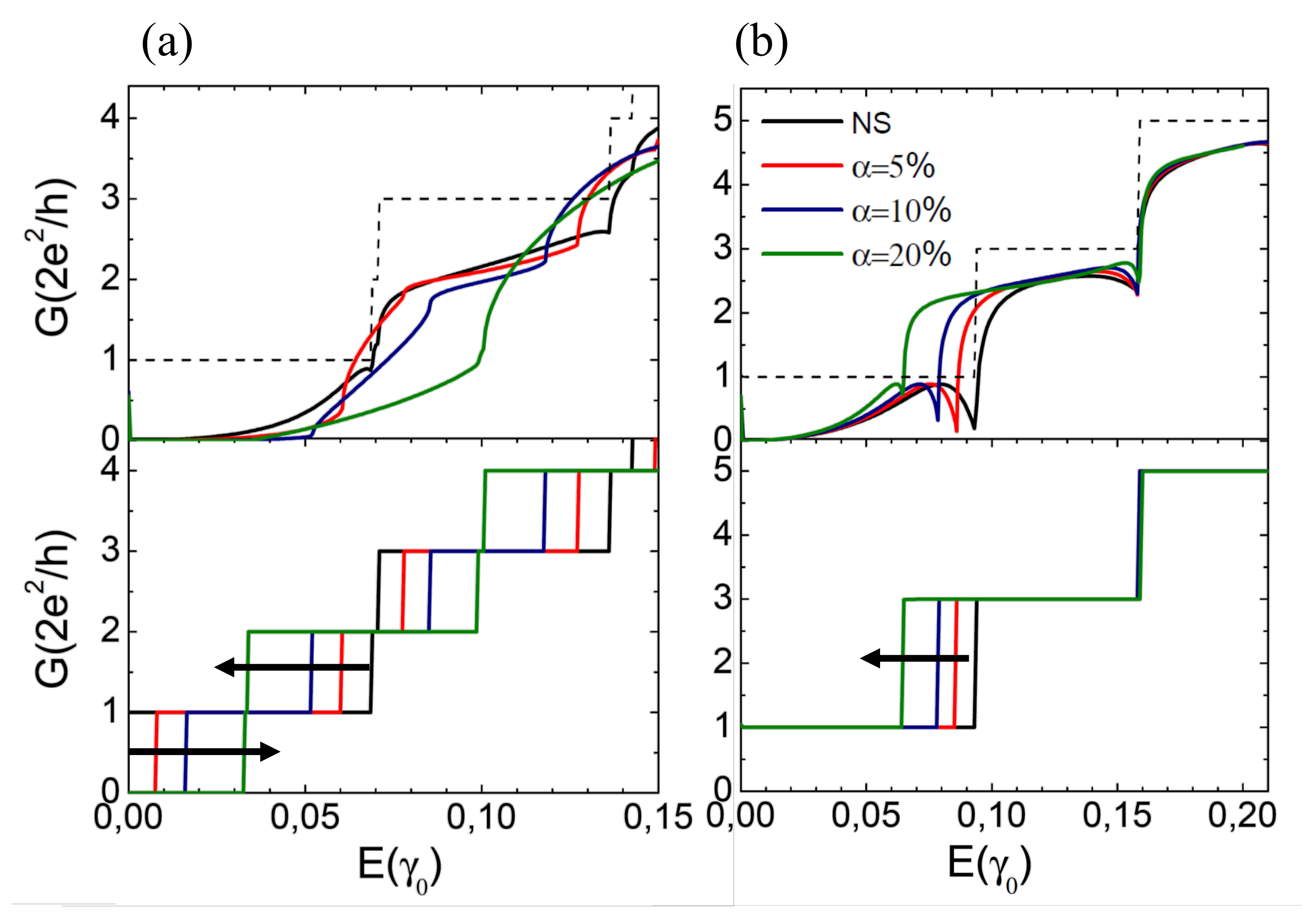}}
\caption{Conductance (top panels) for fold-deformed triangular graphene system with (a) zigzag ($N_Z=39$) and (b) armchair ($N_A=48$) edges, and  $b_f=7.5\;a_{c}$. Dotted lines are the conductance results of the corresponding undeformed leads. Bottom panels: conductance of deformed isolated lead for the different $\alpha$ parameters. \label{foltriam} }
\end{figure}
\end{center}
%%%%%%%%%%%%%%%%%%%%%%%%%%%%%%%%%%%%%%%%%%%%%%%%%%%%%%%%%%%%%%%%%

The electronic transport along fold-like deformed areas have been explored in zigzag nanoribbons \cite{Ramon2016}, revealing that along the stretched region ballistic transport is enhanced in the direction parallel to the deformation. %, with the ribbons acting as a natural waveguides for electronic transport. 
In conductance results, it was observed that as strain increases, the onset of the second conductance plateau moves to lower energies and becomes wider. Similar results remarkably happen for the conductance of the all-folded  armchair hexagonal and triangular flakes with zigzag folded leads studied here. These extra conductance channels are revealed within the energy range corresponding to the first conductance plateau for the undeformed system. 

For all-folded triangular graphene flakes with zigzag edges a clean gap is always opened at energies close to the Fermi energy (Fig.\ref{foltriam}(a)). This may be understood noting that the different conductance plateau of the leads (for instance, first and second) are shifted in opposite directions as a function of the strain intensities, as marked by the  black arrows in the bottom panel of Fig. \ref{foltriam}(a).  The conductance results for the armchair triangle flake system are also guided by the response of the leads to the strain as illustrates in Fig. \ref{foltriam}(b). The general effects of the strain are marked by the main features observed on the transport of the isolated nanoribbons and the geometrical characteristics of the matching edges of the flakes and leads. 

%%%%%%%%%%%%%%%%%%%%%%%%%%%%%%%%%%%%%%%%%%%%%%%%%%%%%%%%%%%%%%%%%
\begin{center}
\begin{figure}[hbt]
\scalebox{0.36}{\includegraphics{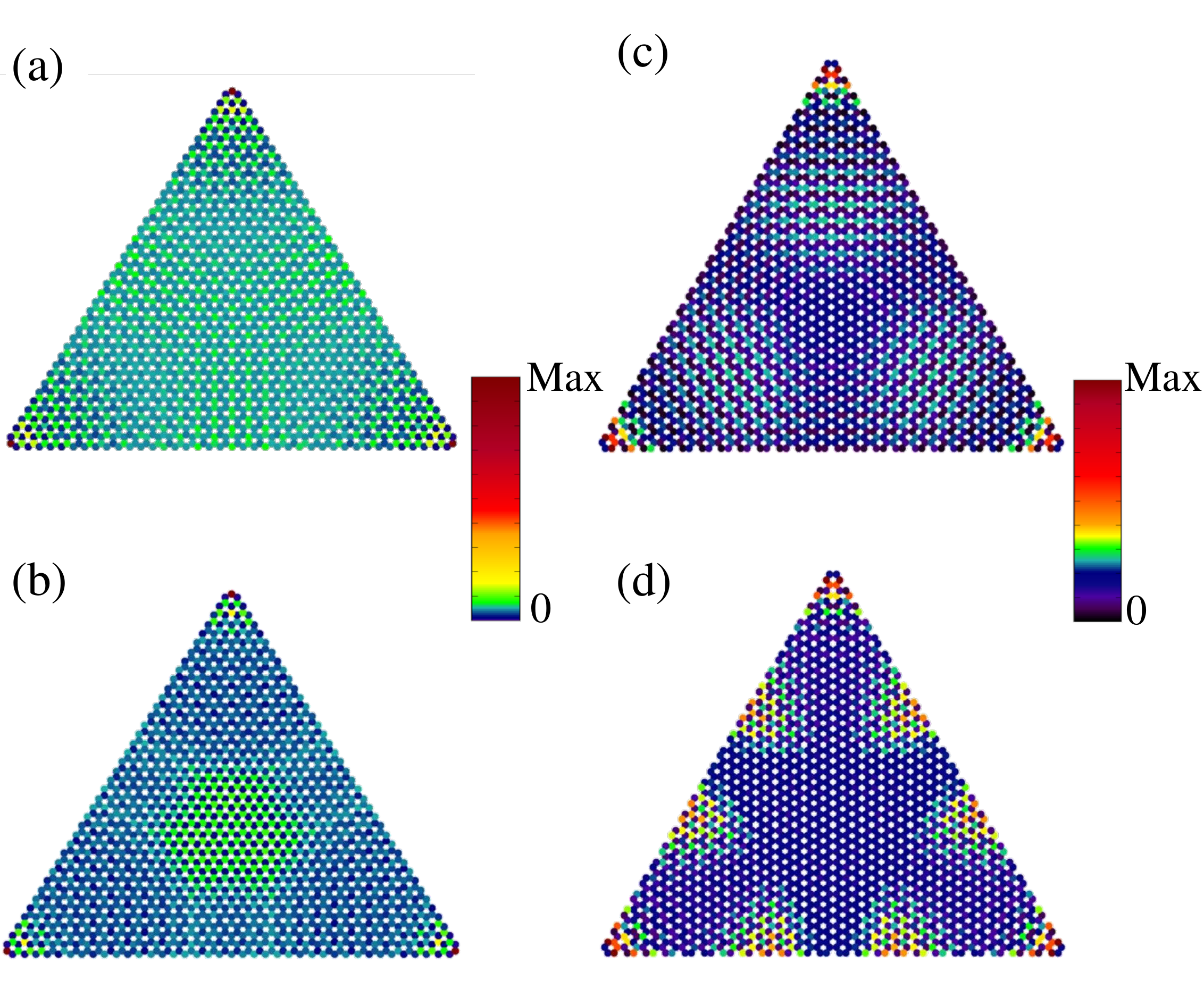}}
\caption{Local electronic density of states for unstrained [(a) and (c)]  and fold deformed [(b) and (d)] triangular-graphene flake systems at $E=0.08\;\gamma_0$. (a) and (b) [(c) and (d)] correspond to zigzag [armchair] edged flakes. The parameter $b_f=7.5\;a_{c}$ is fixed and $\alpha=20\%$. The triangular sizes are the same as used in Fig. 8. {\label{lodstfold} }}
\end{figure}
\end{center}
%%%%%%%%%%%%%%%%%%%%%%%%%%%%%%%%%%%%%%%%%%%%%%%%%%%%%%%%%%%%%%%%%

Local electronic density of state for these triangular graphene systems are shown in Figs. \ref{lodstfold}(a)-(d) for zigzag ($N_Z=39$) and armchair ($N_A=48$) triangular flakes. The results correspond to the case of $\alpha=20\%$ and $b_f=7.5\;a_{c}$, discussed in Figs. \ref{foltriam}(a) and (b). The LDOS of the zigzag triangular flake system  without deformation at the energy $E=0.08\;\gamma_0$ is shown in Fig. \ref{lodstfold}(a). The LDOS presents a homogeneous distribution with lower values at the corners of the flake (except for the extreme carbon atom positions), in contrast with the reported result of isolated triangular flake\cite{Zarenia2011}, when the carriers are confined at the zigzag edges of the triangle leading to localization effects due to the degenerate zero-energy states. The existence of terminals coupled to the central conductor destroys the free boundaries and the corresponding charge accumulation. When the full triangular system is deformed with the folds, the charge accumulation is verified at the central part of the flake leading to a reduced conductance result (see Fig. \ref{foltriam}(a)). This charge distribution follows the respective pseudomagnetic field profile shown in right panel in Fig.\ \ref{device} (c),  that for this geometry is null at the whole system, except at the junction of the three folds.

The other system with armchair edges is coupled with zigzag leads. We remember that extra channels in the central region are generated on fold-deformed zigzag nanoribons\cite{Ramon2016} and they are responsible for the charge concentration at intercalate regions of the lateral sizes of the triangle shown in Fig. \ref{lodstfold}(d).  The evidences of enhanced transport at the same energy of $E=0.08\gamma_0$ when compared to the undeformed system should be expected if we conclude that the leads essentially dictate the electronic transport through the triangle geometry. In graphene nanoribbons, these states forming extra conducting channels at the deformed leads are very peculiar \cite{Ramon2016}. For energies at the second conductance plateau the LDOS shows a high concentration at the deformation region, with sublattice polarization, and formed by states from a single $K$ valley in graphene. States with the same velocity show real space valley polarization, yielding valley-polarized currents moving along different parts of the structure due to the deformation. 

%%%%%%%%%%%%%%%%%%%%%%%%%%%%%%%%%%%%%%%%%%%%%%%%%%%%%%%
\begin{center}
\begin{figure}[hbt]
\scalebox{0.38}{\includegraphics{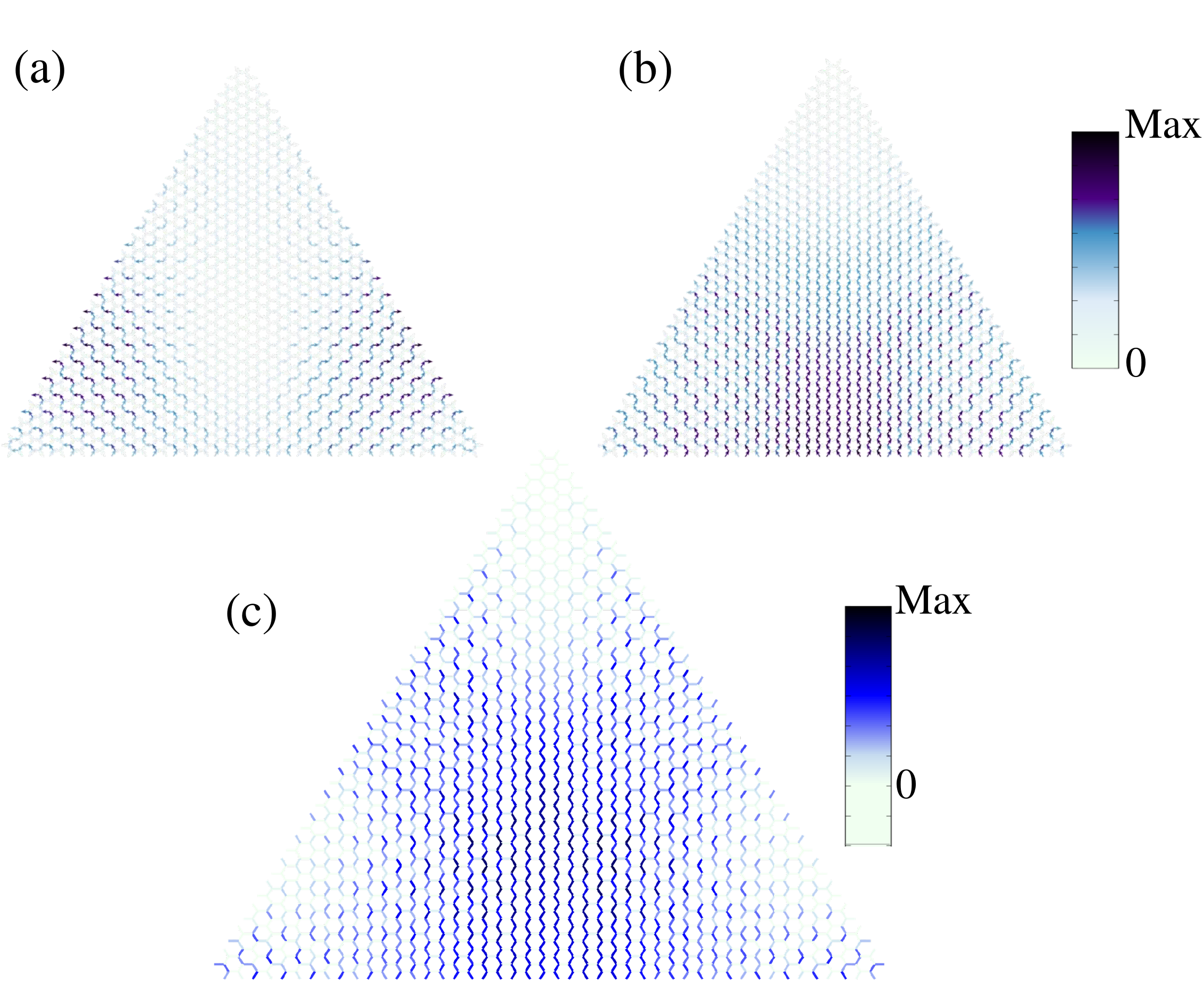}}
\caption{Current density maps of an armchair triangular graphene system (a) without perturbation and (b) with an extended folded strain given by $b_{f}=7.5\;a_{c}$ and $\alpha=20\%$, both calculated at $E=0.08\;\gamma_0$. (c) Variation of the current density maps, $\Delta J=J_f-J_0$. Small arrows in (a) and (b) define the signal of the current density.}{\label{corrientet} }
\end{figure}
\end{center}
%%%%%%%%%%%%%%%%%%%%%%%%%%%%%%%%%%%%%%%%%%%%%%%%%%%%%%%%%%

To further analyze how theses states behave in the multi-terminal system, we explore how the current distribution in the trigonal flake is affected by the folded deformation extended to the leads. The current distribution was calculated by $J_{ij} = Im(T_{ij}A_{ij})/\hbar$, with the spectral function\cite{Mikkel2017} defined as $A_{ij} = (G\Gamma^{L}G^\dagger)_{ij}$.
Fig. \ref{corrientet} shows the real-space density map of a current incident from the bottom lead and passing to the right and left terminals through an armchair triangular flake system at $E=0.08\;\gamma_0$.  From the bottom lead, the current in the unperturbed triangular flake is distributed within a symmetry pattern, decreasing in the central region and upper vertex, as shown in Fig. \ref{corrientet}(a). On the other hand, the triangular fold-deformed flake ($\alpha=20\%$) exhibits an enhanced current density mainly in the flake center [see Fig. \ref{corrientet}(b)]. We also plotted the current density variation, $\Delta J=J_{f}-J_0$, with $J_{f}$ and $J_0$ being the density currents related to the deformed and flat triangular flakes, respectively. This result confirms the increase in the central part of the flake and moreover in the intermediate size at the interfaces with right and left terminals, as depicted in Fig.\ref{corrientet}(c), corroborating with the idea that the current flowing from the bottom lead may be split into the two other leads that works as natural waveguides\cite{Ramon2016} for electronic transport when they are folded deformed. In our calculation for higher energies, we noticed even greater enhancement of the current densities at the central region, that follows a stripped pattern, in agreement with the expected pseudo-magnetic field arising from the the fold deformation.

\begin{center}
\begin{figure}[hbt]
\includegraphics[width=0.95\columnwidth]{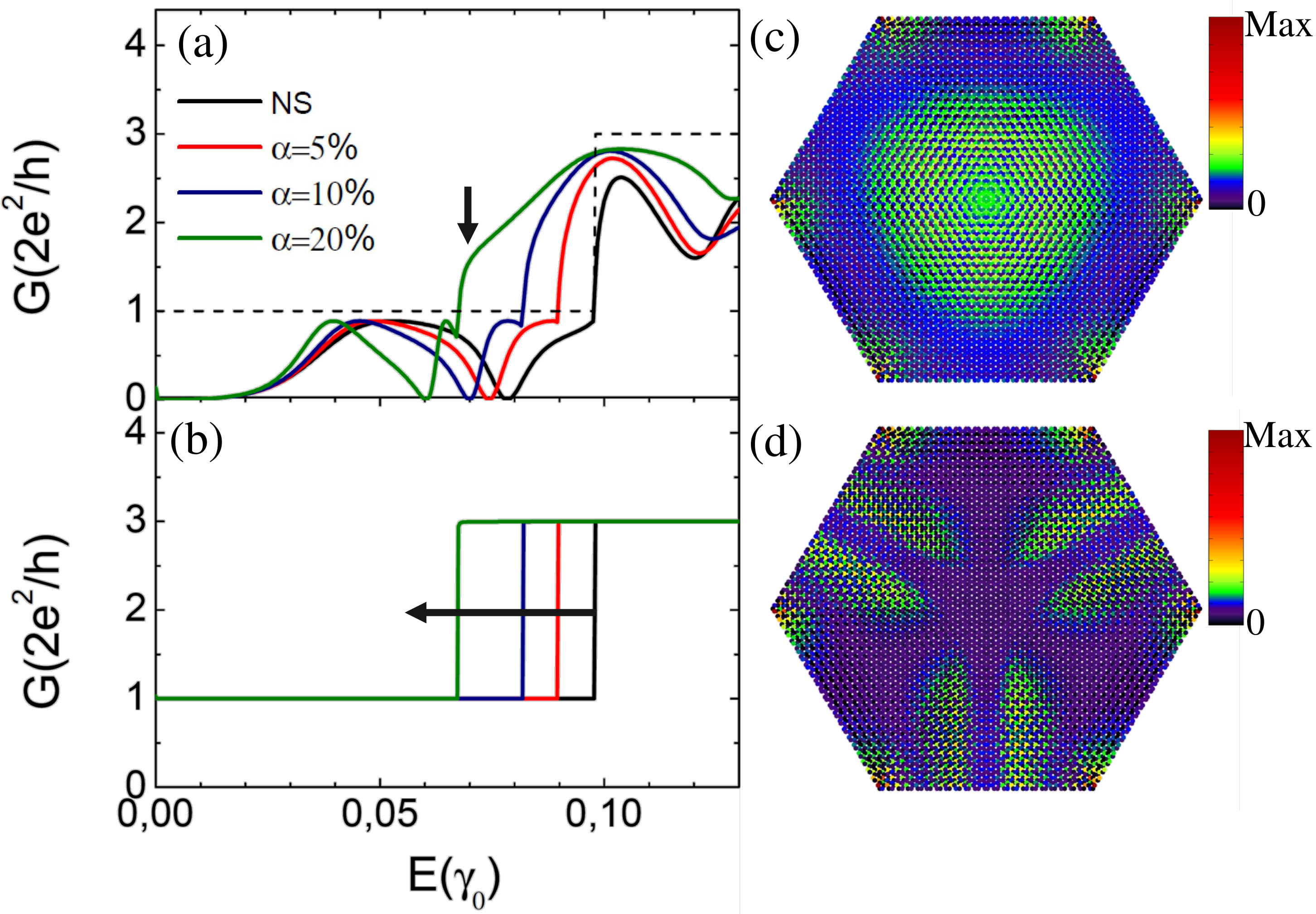}
\caption{(a) Conductance for hexagonal graphene armchair flake ($N_A=50$) with fold-like deformation. The dotted line corresponds to the conductance of the undeformed terminals. (b)  Conductance of the deformed lead for different values of the strain parameter $\alpha$, with fixed $b_f=7.5\;a_{c}$. Local electronic density of states for (c) undeformed  and (d) all-folded hexagonal graphene flake systems with armchair edges at $E = 0.07 \gamma_0$. Strain parameters: $bf = 7.5 a_c$ and $\alpha= 20\%$. \label{foldh} }
\end{figure}
\end{center}

Finally, we also consider the case of the armchair-edged flake (with zigzag leads). Similar features are observed, although the conductance is more suppressed than in the triangular armchair case.  The shift of the conductance plateau of the leads due to changes of the strain intensity, also affects the electronics properties of the graphene flake system, following the general trend of the plateau energy positions shown in the conductance results of the isolated leads as shown in Fig.\ \ref{foldh} (a) and (b).   Results of the electronic local density of states for undeformed and  all-deformed hexagonal multi-terminal system are displayed in  Figs.\ref{foldh} (c) and (d), respectively, at $E = 0.07 \gamma_0$.  Since the deformation is extended in a larger area of the central flake, striped regions with enhanced LDOS concentration can be seen clearly in the flakes, that continue to the leads. In that way a direct path for the electronic current is created enhancing the conductance due to the deformation. Actually, the fold-like deformation extended to the leads reduces possible electronic scattering at the corners of the flake, being possible to turn the orientation of the electronic current with higher conductance in the hexagonal case.

%%%%%%%%%%%%%%%%%%%%%%%%%%%%%%%%%%%%%%%%%%%%%%%%%%%%%%%%%%%%%%%%%
%%%%%%%%%%%%%%%%%%%%%%%%%%%%%%%%%%%%%%%%%%%%%%%%%%%%%%%%%%%%%%%%%

In the case of hexagonal zigzag-edged flakes, the fold-like deformation also affects the conductance promoting shifts in the typical resonant states (not shown here). The change in the energy of  the resonant peaks follow the conductance evolution of the folded armchair leads (presented in the bottom part of Fig. \ref{foltriam}(a)).

\section{Conclusions}
  
We have shown that multi-terminal systems with deformations can be used to obtain desired transport properties in the graphene-flake devices, by changing the deformations parameters, the geometry and the system edge type. In the case of non extended deformations, we observed that central flakes with more atoms, i. e., bigger area, favor resonant states at the deformation region. In particular, for circular Gaussian bumps and smoothed-fold deformations, the hexagonal graphene flake would be the most promise system studied here to explore the deformation effects. On the other hand, for the three-fold extended deformations investigated, since the perturbations are also present at the leads, the triangular graphene flake presents a better transport response due to reduction of electronic scattering at the central part of the system. In fact, the three-fold deformation for triangular armchair graphene flakes acts as a waveguide for the current, favoring transport to the designed terminals, following the zigzag directions. Similar effects are also observed for hexagonal armchair graphene flakes, although the conductance is more affected by scattering at the central part of the system. For triangular graphene flakes with both terminations, it is possible to control the system gap just by changing the deformation parameters. These findings could be tested in transport measurements in properly prepared substrates. The necessary strain intensities considered in our calculations can be achieved in current experimental settings. 

Furthermore, it has been reported that the deformations explored in this work (folds and  Gaussian bumps), and other deformations such as triaxial in-plane strain, could produce filters\cite{Ramon2016} and beam splitters\cite{Mikkel2017}, for obtaining valley polarized currents. These valley polarized currents are a natural consequence of the non-equivalent pseudo-magnetic field for states from both valleys in graphene, and could also be explored in multi-terminal graphene flakes. In particular, our discussion on three-folded graphene flakes has shown that it is possible tuning current density routes in the real space. Therefore, it could be an interesting scenario for building nanoscale waveguides for valley polarized currents.

\section{Acknowledgments}

We  thank Nancy Sandler, Eva Y. Andrei, and R. Carrilo-Bastos for interesting discussions. This work has been financially supported by  FAPERJ under grant E-26/102.272/2013 and /202.953/2016. We acknowledge the financial support of the CNPq and the INCT em Nanomateriais de Carbono. VT would like to thanks FAPESP under grant 2017/12747-4.

\end{document}